\documentclass[structabstract]{aa} 
\usepackage{graphicx}
\usepackage{txfonts}
\usepackage{url}
\usepackage[round]{natbib}
\usepackage{natbib}
\usepackage{float}

\bibpunct{(}{)}{;}{a}{}{,}
\DeclareMathSymbol{\lesssim}{\mathrel}{AMSa}{"2E}
\def\ltsima{$\; \buildrel < \over \sim \;$}
\def\lsim{\lower.5ex\hbox{\ltsima}}
\def\gtsima{$\; \buildrel > \over \sim \;$}
\def\gsim{\lower.5ex\hbox{\gtsima}}
\begin{document}
\bibliographystyle{aa}
 \title{1RXS J180408.9-342058: an ultra compact X-ray binary candidate with a transient jet\thanks{Based on observations made with ESO Telescopes at the La Silla Observatory under programme ID 094.D-0692(B)}}

   \author{M. C. Baglio\inst{1, 2},
          P. D'Avanzo\inst{2},
          S. Campana \inst{2},
          P. Goldoni \inst{3},
          N. Masetti \inst{4, 5},
          T. Mu\~{n}oz-Darias\inst{6, 7},
          V. Pati\~{n}o-\'{A}lvarez\inst{8}
          \and
          V. Chavushyan\inst{8}
          }
		  
   \institute{Universit\`{a} dell'Insubria, Dipartimento di Scienza e Alta Tecnologia, Via Valleggio 11, I–22100 Como, Italy                      \\
              \email{cristina.baglio@brera.inaf.it}
         \
             \and 
             INAF, Osservatorio Astronomico di Brera, Via E. Bianchi 46, I-23807 Merate (LC), Italy
             \and
             APC, Astroparticules et Cosmologie, Universit\'{e} Paris Diderot, CNRS/IN2P3, CEA/Irfu, Observatoire de Paris, Sorbonne Paris Cit\'{e}, 10, rue Alice Domon et L\'{e}onie Duquet, F-75205 Paris Cedex 13, France
             \and
             INAF-Istituto di Astrofisica Spaziale e Fisica Cosmica di Bologna, Via Gobetti 101, I-40129 Bologna, Italy
             \and
             Departamento de Ciencias F\'{i}sicas, Universidad Andr\'{e}s Bello,
  Fern\'{a}ndez Concha 700, Las Condes, Santiago, Chile
  \and
             Instituto de Astrofisica de Canarias, E-38205 La Laguna, Tenerife, Spain       
             \and
             Departamento de astrofisica, Universidad de La Laguna, E-38206 La Laguna, Tenerife, Spain
             \and
             Instituto Nacional de Astrof\'{i}sica, \'{O}ptica y Electr\'{o}nica, Apartado Postal 51-216, 72000 Puebla, M\'{e}xico
              }          
   \date{ }

   \abstract
   {}
   {We present a detailed NIR/optical/UV study of the transient low mass X-ray binary 1RXS J180408.9-342058 performed during its 2015 outburst, aimed at determining the nature of its companion star. }
   {We obtained three optical spectra ($R\sim 1,000$) at the 2.1 m San Pedro M\'{a}rtir Observatory telescope (M\'{e}xico). We performed optical and NIR photometric observations with both the REM telescope and the New Technology Telescope (NTT) in La Silla. We obtained optical and UV observations from the \textit{Swift} archive. Finally, we performed optical polarimetry of the source by using the EFOSC2 instrument mounted on the NTT.}
   {The optical spectrum of the source is almost featureless since the hydrogen and He I emissions lines, typically observed in LMXBs, are not detected. Similarly, carbon and oxygen lines are neither observed. We marginally detect the He II 4686 $\AA$ emission line, suggesting the presence of helium in the accretion disc. No significant optical polarisation level was observed.}
   {The lack of hydrogen and He I emission lines in the spectrum implies that the companion is likely not a main sequence star. Driven by the tentative detection of the He II 4686 $\AA$ emission line, we suggest that the system could harbour a helium white dwarf. If this is the case, 1RXS J180408.9-342058 would be an ultra-compact X--ray binary. By combining an estimate of the mass accretion rate together with evolutionary tracks for a He white dwarf, we obtain a tentative orbital period of $\sim$40 min. 
On the other hand, we also built the NIR--optical--UV spectral energy distribution (SED) of the source at two different epochs. One SED was gathered when the source was in the soft X--ray state, and it is consistent with the presence of a single thermal component. The second SED, obtained when the source was in the hard X--ray state, shows a thermal component together with a tail in the NIR, likely indicating the presence of a (transient) jet.}

\authorrunning{Baglio, M. C. \& al.} 
\titlerunning{1RXS J180408.9-342058: an ultra compact X-ray binary candidate with a transient jet}
\maketitle

\section{Introduction}
Low mass X--ray binaries (LMXBs) are accreting binary systems hosting a compact object (either a stellar massive black hole or a neutron star) and a companion star that is usually a main sequence, low-mass star. In such systems the transfer of mass takes place via Roche lobe overflow and the consequent formation of an accretion disc around the compact object. LMXBs usually have orbital periods varying from minutes to days, depending on the nature of the companion star. In particular, it exists a class of tight LMXBs, called ultra compact X-ray binaries (UCXBs), that are characterised by orbital periods $<$ 80 min, and usually harbour white dwarfs or hydrogen-poor stars as companions.

LMXBs can be both transient or persistent. The former group alternate short (weeks--months) and sudden periods of outburst, characterised by a typical X-ray luminosity of $ 10^{36}-10^{38} \, \rm erg/s $, during which the compact object accretes at high rates, to long lasting (years to decades) periods of quiescence, when the X-ray luminosity drops up to seven orders of magnitude. 

\begin{table*}[htb]
\caption{Complete log and results of the NTT observations performed on Feb. 8, 2015. The magnitude values (Vega) are not corrected for reddening, which is reported in the 6th column \citep{Schlafly2011}. In the last column, the $ \chi^2/\rm dof $ of the constant fits performed on the light curves are reported. Errors are indicated at a $90 \%$ c.l. }            
\label{NTT_log}      
\centering                       
\begin{tabular}{c c c c c c c c c }       
\hline\hline                
Filter  & Exposures& UT mid observation  & Wavelength  & Mean  & $A_{\lambda}$ & $\chi^2/\rm dof$\\   
        &        (s)           & YYYYmmdd (MJD)             & (\AA) &   Magnitude              & (mag) 	& \\
\hline                     
   $B$ &$8\times15$  & 20150208.3735 (57061.3735)&4400& $18.03 \pm 0.03$ & $1.73 \pm 0.12 $ & 15.41/4\\
   $V$ &$8\times15$  & 20150208.3708 (57061.3708)&5476 & $17.71 \pm 0.02$ & $1.30 \pm 0.09 $ & 52.63/7\\
   $R$ &$8\times15$  &20150208.3614 (57061.3614)&6431 & $ 17.24 \pm 0.03$ &$1.04 \pm 0.07$  & 2.7/4 \\
   $i$ &$8\times15$    &20150208.3661  (57061.3661)&7931& $17.03 \pm 0.06$ & $0.75 \pm 0.05$& 6.389/7 \\
\hline                             
\end{tabular}
\end{table*}

\begin{table*}[htb]
\caption{Complete log and results of the REM observations in the two epochs (Feb. 26, 2015 and Apr. 24, 2015).
The magnitude values (AB magnitudes) are not corrected for reddening, whose parameters are in the 6th 
column \citep{Schlafly2011}. In the last column, the $ \chi^2/\rm dof $ of the constant fits performed (where possible) on the light curves are reported. Errors are indicated at a $90 \%$ c.l. }          
\label{obs_log}      
\centering                          
\begin{tabular}{c c c c c c c }       
\hline\hline                 
Filter  & Exposures & UT mid observation  & Wavelength & Mean  & $A_{\lambda}$ & $\chi^2/\rm dof$\\   
         &       (s)            & YYYYmmdd (MJD)                & (\AA) 	        &   Magnitude              & (mag) 	& \\ 
\hline
\multicolumn{7}{|c|}{2015 February 26}\\
\hline                       
   $g$ &$32\times30$   &20150226.3423 (57079.3423) &4770& $17.84 \pm 0.05$ & $1.56 \pm 0.06 $ & 58.3/30\\
   $r$ &$32\times30$  &20150226.3423 (57079.3423)  &6231 & $17.38 \pm 0.03$ & $1.06 \pm 0.04 $ & 78.4/27\\
   $i$ &$32\times30$   &20150226.3423 (57079.3423) &7625 & $17.02\pm 0.03$ &$0.75 \pm 0.03$  & 42.9/31 \\%
   $J$ &$22\times15$  &20150226.3394 (57079.3394) &12350& $15.79 \pm 0.12$ & $0.33 \pm 0.01 $ & -- \\
   $K$ &$4\times15$   &20150226.3309 (57079.3309) & 21590& $14.68 \pm 0.26$ & $0.14 \pm 0.01 $ & --\\
\hline                                  
\multicolumn{7}{|c|}{2015 April 24}\\
\hline   
  $g$ &$12\times30$  &20150424.2584 (57136.2584) &4770& $17.02 \pm 0.04$ & $1.56 \pm 0.06 $ & 5.3/11\\
   $r$ &$12\times30$  &20150424.2584 (57136.2584) &6231& $16.78 \pm 0.03$ & $1.06 \pm 0.04 $ & 12.9/11 \\
   $i$ &$12\times30$  &20150424.2584  (57136.2584) &7625& $ 17.08\pm 0.03$ &$0.75 \pm 0.03$  & 28.2/11 \\
   $J$ &$8\times15$  &20150424.2437 (57136.2437) & 12350 & $16.84 \pm 0.16$ & $0.33 \pm 0.01 $ & -- \\
\hline                                  
\end{tabular}
\end{table*}

The X-ray source 1RXS J180408.9-342058 (hereafter J180408) was first identified as a neutron star (NS) LMXB after the INTEGRAL detection of a type I thermonuclear burst in 2012 \citep{Chenevez2012}. Assuming the Eddington luminosity as an upper limit, they limited its distance to 5.8 kpc. Since then, the source remained in quiescence until Jan. 22, 2015, when the \textit{Swift}/BAT detected a new outburst \citep{Krimm2015a}. This outburst was characterised by a slow rise without the detection of Type I bursts. A MAXI/GSC X-ray spectrum on Jan. 26-28, 2015 showed a hard spectrum, that can be modelled by a power--law with index $ 1.68 \pm 0.27 $ \citep{Negoro2015}. Since the detection of the outburst, the source was regularly monitored by \textit{Swift}. In particular, \citet{Krimm2015b} reported an increase in brightness in the hard X-ray band (15-50 keV) after the first 20 days of activity, that then plateaued at an average rate of $ \sim 100 $ mCrab. 
Radio observations obtained with the VLA, together with contemporaneous \textit{Swift} X-ray observations showed that the position of the source in the radio/X-ray luminosity plane was consistent with that of hard-state neutron star LMXBs \citep{Deller2015}. While the source was in its hard state, radio emission consistent with jets were also detected in the radio band (\citealt{Deller2015}; \citealt{Degenaar2015}).

Around Apr. 3, 2015, a significant drop in the hard X-ray flux was detected, together with an increase in the softer X-rays (2-10 keV), suggesting that the source transitioned to a soft X-ray spectral state \citep{Degenaar2015}. 

\section{Observation and data analysis}

\subsection{Optical spectroscopy}
J180408 was observed on Apr. 20, 2015 with the 2.1 m telescope at the San Pedro M\'{a}rtir Observatory (M\'{e}xico), equipped with the Boller \& Chivens spectrograph. Three 1,800 s spectra with a resolution of $ \sim 6.5\, \AA $ (350 $\rm km\, \rm s^{-1} $) were acquired, covering the wavelength range 4000--7800 \AA. The three spectra were combined and averaged to obtain one final spectrum, in order to increase the signal to noise (S/N) ratio. 

Data reduction was carried out using standard procedures for both bias subtraction and flat-field correction using IRAF\footnote{http://iraf.noao.edu/}. The wavelength calibration was done using neon-helium-copper-argon lamps. The flux calibration was performed against the catalogued spectroscopic standard star Feige 67 \citep{Massey1988}.

\subsection{NTT photometry}
The LMXB J180408 was observed at the La Silla observatory with the New Technology Telescope (NTT) on Feb. 8, 2015 using the instrument EFOSC2 equipped with the optical Bessel $BVR$ and Gunn $i$ filters ($ 4400\AA -7930 \AA $). 
A total of $8\times15\,\rm s$ images for each filter was obtained in each band. A log of the observations is reported in Tab. \ref{NTT_log}.

Bias and flat field corrections have been applied using the standard procedures (subtraction of an average bias frame and division by a normalised average flat field). Due to the crowdedness of the field, all the magnitudes were extracted performing point spread function (PSF) photometry with the ESO MIDAS\footnote{http://www.eso.org/sci/software/esomidas//} {\tt daophot} task \citep{Stetson1987}. 
The night was clear, with seeing $ \sim 0.7-0.8'' $ for the whole duration of the observation. We performed differential photometry with respect to a selection of fourteen isolated stars in the field.

\begin{table*}[htb]
\caption{Complete log and results of the \textit{Swift/UVOT} observations taken on Feb. 26, 2015. The (Vega) magnitudes are not corrected for reddening, whose parameters are reported in the last coloumn \citep{Schlafly2011}. Errors are indicated at a $90 \%$ c.l. }             
\label{tab_log_UVOT}      
\centering                         
\begin{tabular}{c c c c c c}       
\hline\hline                
Filter  & Exposure time & UT mid observation & Wavelength & Magnitude  & $A_{\lambda}$  \\    
         &          (s)              &     YYYYmmdd (MJD)& ($\AA$) 	        &                & (mag) 	 \\ 
\hline                        
   $b$ &122   &20150226.1099 (57079.1099) & 4392& $17.22\pm 0.08$ & $1.66 \pm 0.06 $ \\
   $u$ &122   &20150226.1088 (57079.1088) & 3465& $16.45 \pm 0.06$ & $2.01 \pm 0.07 $ \\
   $w1$ &244 & 20150226.2133 (57079.2133) & 2600& $16.90 \pm 0.07$ & $2.69 \pm 0.10 $ \\
   $m2$ &322 &20150226.1203 (57079.1203) & 2246& $17.86 \pm 0.12$ & $3.73 \pm 0.14 $ \\
   $w2$ &488 &20150226.1140 (57079.1140) & 1928& $17.58 \pm 0.08$ & $3.31 \pm 0.12 $ \\
\hline                                 
\end{tabular}

\noindent $v$ data were not used due the simultaneous occurrence of a Type I burst.
\end{table*}

\begin{table*}[htb]
\caption{Complete log and results ( 3$ \sigma $ upper limits) of the NTT polarimetric observations taken on Feb. 9, 2015. }         
\label{log_pol}      
\centering                      
\begin{tabular}{c c c c}        
\hline\hline             
Filter  & Exposure time       & UT mid observation   & 3$ \sigma $ upper limit\\    
         &  per pos. angle  (s)  &     YYYYmmdd (MJD)& \\
\hline                        
   $B$ &160 & 20150209.3670 (57062.3670) & $<6.9\%$\\
   $V$ &120 & 20150209.3589 (57062.3589) & $ <5.6\% $\\
   $R$ &120 & 20150209.3414 (57062.3414) & $ <1.4\% $\\
   $i$ &180  & 20150209.3498 (57062.3498) & $<1.6\%$\\
\hline                              
\end{tabular}
\end{table*}

\subsection{REM observations}\label{REM_red}
J180408 was observed in the optical (SDSS $griz$ filters simultaneously) and NIR (2MASS $JK$ bands, one filter at a time) with the REM telescope (\citealt{Zerbi2001}; \citealt{Covino2004}) 
on 2015 Feb. 26, 2015 and Apr. 24, 2015, i.e. before and soon after the change of its X-ray state from hard to soft \citep{Degenaar2015}. 
A complete log of the observations is reported in Tab. \ref{obs_log}.

Each image was bias and flat--field corrected using standard procedures, and PSF photometry techniques were applied in order to obtain the magnitudes of all the objects in the field.
Both nights were clear and seeing remained constant during the whole duration of the observations. We performed differential photometry with respect to a selection of the same six bright and isolated stars of the field in order to correct for any instrumental effect.
The target is well detected in the $gri$ filters, but not in the $z$-band, that was excluded from the analysis. 

The calibration of the optical photometry (NTT and REM) was performed through the comparison between the fluxes measured for a group of isolated stars in the fields, chosen as reference, with the fluxes tabulated in the APASS\footnote{http://www.aavso.org/download-apass-data} catalogue (Sec. \ref{REM_red}). For the NTT observations, in order to pass from the $gri$ magnitudes (i.e. the SDSS filters used in the APASS catalogue) to the $BVR$ magnitudes (Johnson filters, used in EFOSC2) the corrections reported in \cite{Jordi2006} were applied.

\subsection{\textit{Swift/UVOT} observations}\label{sec_UVOT}
After the onset of the outburst, the system J180408 has been extensively observed both in the X-rays and UV frequencies by the \textit{Swift} satellite. In particular, we analysed the UVOT data from the observations that started on 2015 Feb. 26, 2015, because this was contemporaneous to the first set of REM observations. A complete log of the UVOT observations is reported in Tab. \ref{tab_log_UVOT}.

The analysis of the data was performed using the HEASOFT routine \textit{uvotsource} that, once the extraction region (a circle centred on the source, with a radius of 4 arcsec) and the background (a circle with radius of 9.5 arcsec) are defined.

\subsection{Optical polarimetry with NTT}
J180408 was observed on 2015 Feb. 9, 2015 with the optical polarimeter of EFOSC2, mounted on the NTT, equipped with the $ BVRi $ filters. The night was clear (seeing $ \sim 0.7'' $ during the whole duration of the observation). Image reduction was carried out by bias-subtraction and division by a normalised flat-field. A complete log of the observation is presented in Tab. \ref{log_pol}.
All the flux measurements have been performed using the technique of PSF photometry (as explained in Sec. \ref{NTT_obs}). 


A Wollaston prism was inserted in the optical path, so that incident radiation was simultaneously split into two different beams with orthogonal polarisation (ordinary and extraordinary beams, $o$- and $e$- beams). Furthermore, a Wollaston mask prevented the possible overlap of the different images.
The instrument made use of a rotating half wave plate (HWP) in order to take images at four different angles with respect to the telescope axis ($ \Phi_{i} = 22.5^{\circ}(i-1)$, $i=0,1,2,3$). One image of 15 s integration was obtained for each HWP angle for the $ BVRi $ filters. 

\section{Results}

\begin{figure*}
\centering
\includegraphics[scale=0.5]{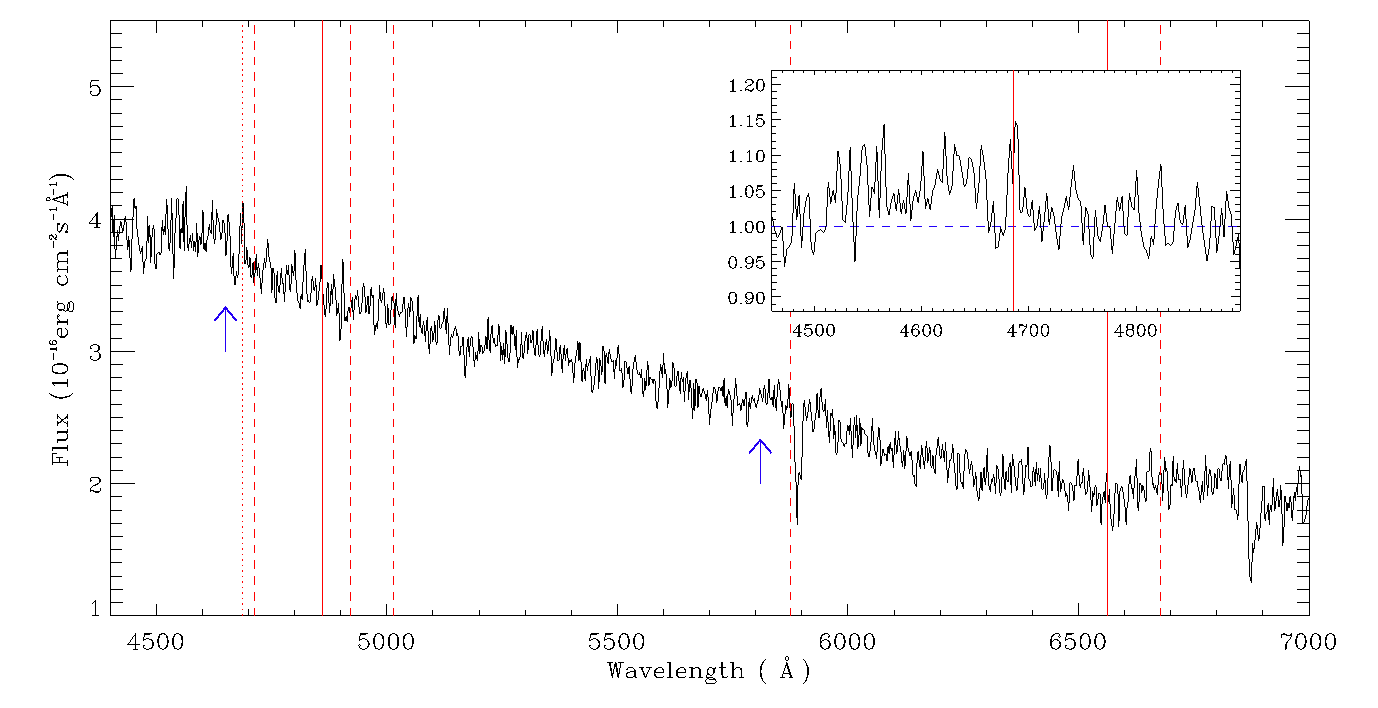}
\caption{Averaged, flux-calibrated spectrum of J180408 obtained at the San Pedro M\'{a}rtir Observatory (Mexico) on Apr. 20, 2015 (MJD 57132). The spectrum is not corrected for reddening. With solid lines, the positions of the most common hydrogen lines are highlighted ($ H_{\alpha}= $6563 \AA\, and $ H_{\beta} =$4861 \AA); with dashed lines are indicated the most prominent He I emission line positions (4713, 4921, 5015, 5876 and 6678 \AA); finally, with a dotted line we indicated the wavelength of the He II line at 4686 \AA. Two blue arrows indicate the tabulated position of the two suggested carbon lines (C II 4650 $\AA$ and C III 5810 $\AA$).
\textit{Inset}: A zoom of the spectrum centered on the possibly detected He II line after the normalisation of the spectrum continuum to 1 is represented. With a solid line, the position of the He II line (4686 \AA) is highlighted.}
\label{spectrum}
\end{figure*}

\subsection{Optical spectroscopy}\label{sec_spectrum}
The flux-calibrated, barycentric corrected and co-added spectrum is shown in Fig. \ref{spectrum}.  Absorption features due to the Galactic sodium doublet 
at 5890 \AA\, and to the MgIb band at 5175 \AA \, were detected.
The spectrum does not show prominent emission lines at variance with LMXBs in outburst. We indicated in Fig. \ref{spectrum} the most common hydrogen emission lines  ($ H_{\alpha} $ at 6563 \AA \, and $ H_{\beta} $ at 4861 \AA), whereas with dashed lines the positions of the most prominent He I emission lines are shown (4713, 4921, 5015, 5876 and 6678 \AA) and none is detected.
On the contrary, we observe the presence of an emission feature at $ \lambda=4686.3 \pm 1.5 \, \AA$, that could correspond to the 4686 \AA \, He II (see dotted line in Fig. \ref{spectrum}). 
According to an F-test, the fit of this line with a Gaussian function is preferable to that with a constant at a $ \sim 4 \sigma$ significance over the 4670--4705\,\AA\ range). This seems to suggest that a certain quantity of Helium may be present in the accretion disc of the system. 
For this possible emission line we derived an equivalent width (EW) of $ 1.5\pm 0.2 \AA $ and a systemic velocity of $19.2 \pm 96.1 \,\,\rm km/s$.

\subsection{NIR/Optical/UV photometry}

\subsubsection{NTT optical observations}\label{NTT_obs}
The results of the NTT $BVRi$ photometry performed on Feb. 8, 2015 (MJD 57061) are reported in Tab. \ref{NTT_log}.   
The optical counterpart of J180408 is well detected in each image (Bessel $ BVR $ and Gunn $ i $). The source displays a clear intrinsic variability in all bands, even on a relatively short timescale (minutes). As shown in Tab. \ref{NTT_log}, the fit of the light curves with a constant function gives as a result values of $ \chi^2/\rm dof $ that prove the low significance of the fit, due to the high erratic variability of the source. 

\begin{figure}[!h]
\begin{center}
\includegraphics[scale=0.46]{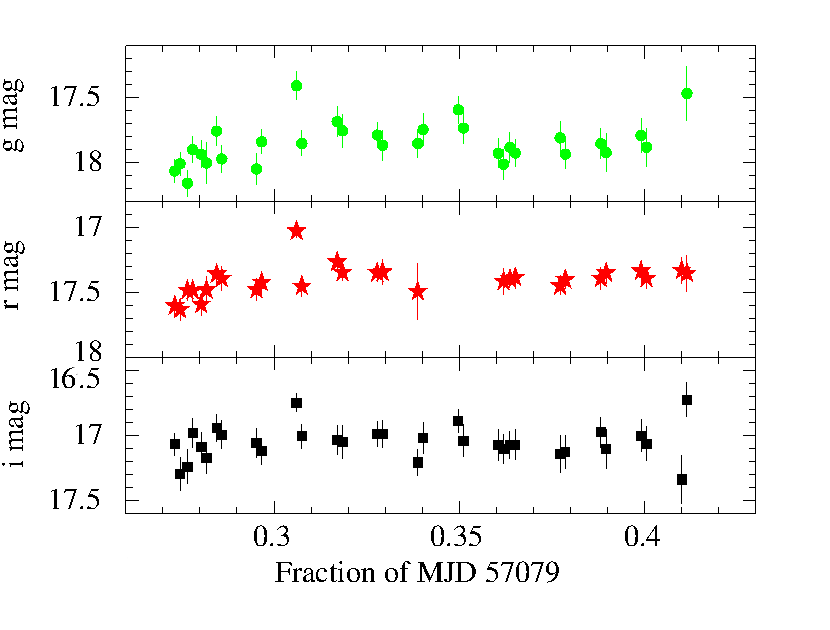}
\caption{From top to bottom, $g$, $r$, $i$ light curves for the system J180408 observed with REM on Feb. 26, 2015. Errors are indicated at the $68\%$ confidence level.}   
\label{fig_REM_feb}
\end{center}
\end{figure}

\begin{figure}
\includegraphics[scale=0.46]{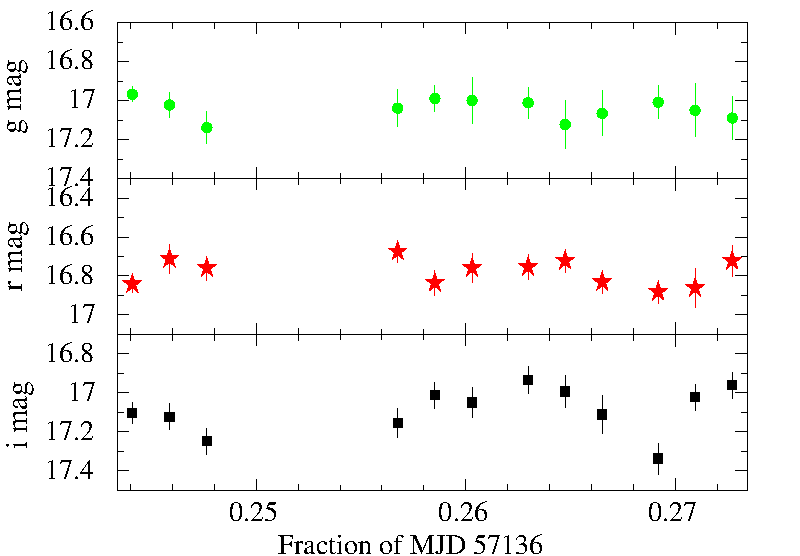}
\caption{From top to bottom, $  gri$ light curves obtained with the REM telescope on Apr. 24, 2015. Errors are indicated at the $ 68\%$ confidence level.}
\label{lc_REM_apr}
\end{figure}

\subsubsection{REM observations}
\label{REM_sec}

The $ gri $ light curves of J180408 obtained with the REM telescope on 2015 Feb. 26, 2015 (MJD 57079, before the state change) are shown in Fig. \ref{fig_REM_feb}.
The source shows intrinsic variability even at this epoch of observation, resulting in high values of $\chi^2/\rm dof$ in case of a constant fit (Tab. \ref{obs_log}). 
Similar conclusions can be deducted from the light curves obtained 
for J180408 on 2015 Apr. 24, 2015 (MJD 57136, Fig. \ref{lc_REM_apr}), i.e. soon after the transition from the hard to the soft X-ray state. 
The mean magnitudes obtained in the $ g,r,i $-bands from the fit of the light curves with a constant are reported in Tab. \ref{obs_log}. 



In the NIR, differently from the optical observations, we do not possess enough data to build light curves. Thus, we averaged all the images to derive the infrared magnitudes reported in Tab. \ref{obs_log}.

\subsubsection{\textit{Swift/UVOT} observations}
J180408 was observed by the \textit{Swift} satellite on Feb. 26, 2015 (MJD 57079, Tab. \ref{tab_log_UVOT}). 
\textit{Swift}/UVOT observed with all its filters, however due to the contemporaneous detection of a Type-I burst, the $ v $-band image has been excluded from the analysis. The calibrated magnitudes of the target (obtained as described in Sec. \ref{sec_UVOT}), together with the absorption coefficients for each filter \citep{Schlafly2011} are reported in Tab. \ref{tab_log_UVOT}.

\subsection{Optical polarimetry with the NTT}\label{Opt_pola}

With the aim of evaluating the degree of linear polarisation $ P $ of the system, we computed, for each value of the instrumental position angle $ \Phi $, the quantity:
\begin{equation}\label{eq_S_par}
S(\Phi)=\left(\frac{f^{o}(\Phi)/f^{e}(\Phi)}{\left\langle f^{o}_{u}(\Phi)/f^{e}_{u}(\Phi_{i})\right\rangle  }-1\right)/\left(\frac{f^{o}(\Phi)/f^{e}(\Phi)}{\left\langle f^{o}_{u}(\Phi)/f^{e}_{u}(\Phi)\right\rangle  }+1\right),
\end{equation}

where $ f^{o} $, $ f^{e} $ are the fluxes in the ordinary and extraordinary beams and 
$ \left\langle f^{o}_{u}(\Phi)/f^{e}_{u}(\Phi)\right\rangle  $ is the averaged ratio between the ordinary and extraordinary fluxes of the unpolarised field stars, chosen as reference.
As exhaustively described in \citet{Serego}, this parameter gives an estimate of the projection of the polarisation along the different directions. 
The parameters $ S $ and the polarisation degree and angle ($ P $ and $ \theta_{\rm P} $, respectively) are then related through the relation:

\begin{equation}\label{fit_cos}
S(\Phi)= P\cos 2\left( \theta -\Phi\right).
\end{equation}

The fit of $ S(\Phi) $ with Eq. \ref{fit_cos} returns $ P $ from the semi-amplitude of the oscillation and $\theta_{\rm P}$ from the position of the first maximum of the curve.
This method makes any additional correction (interstellar and instrumental effects) unnecessary, since $ S $ is already normalised to the non-polarised reference stars.
The polarisation degrees are consistent with zero in all bands.
We derived 3$ \sigma $ upper limits to the optical polarisation of the target, that are quoted in Tab. \ref{log_pol}.
Except for the $ B $- and $ V $- bands, the upper limits obtained are lower than the expected maximum interstellar contribution to the LP of $ 3.9\% $, evaluated following \citet{Serkowski75}.


\section{Discussion}

\subsection{A possible ultra compact nature for J180408}
The apparent absence of H and He I lines is uncommon for typical LMXBs that harbour a main sequence companion star. 
LMXBs may also host a degenerate companion, like a white dwarf or a hydrogen-poor star. In this case to fill the Roche lobe the orbital period should be short and these LMXBs belong to the UCXBs class.
\citet{Nelemans2004} analysed the spectra of three UCXBs (4U 0614+091, 4U 1543-624, and 2S 0918-549). In none of them hydrogen and He I emission lines were detected, as for J180408. The case of 4U 0614+091 was particularly intriguing since its spectrum shows prominent emission lines that could not be connected to hydrogen and helium lines. Instead, they were consistently identified with carbon and oxygen lines, leading to the conclusion that the companion star was a carbon-oxygen (CO) WD. Similar conclusions were also deducted for the other UCXBs reported in \citet{Nelemans2004}.

As described in Sec. \ref{sec_spectrum}, the spectrum of J180408 is rather featureless, except for the possible detection of a weak He II 4686$\AA$ line, suggesting that helium is present in the accretion disc of the system. However, the total absence of H and He I lines is unusual and reminiscent of the behaviour of UCXBs. We thus compared the spectrum of 4U 0614+091 with that of J180408 aiming at identifying possible carbon and oxygen emission lines in the spectrum of our target. 
Despite the moderate S/N of the spectrum (Fig. \ref{spectrum}), no clear evidence for C/O lines can be derived. Just the hints of two possible emission lines corresponding to the ones listed in \cite{Nelemans2004} and \cite{Baglio_4U0614_2014} are found: the C II 4650 \AA \, and the C III 5810 \AA \, lines.
The C II 4650 \AA \, line may be confused with the so-called Bowen blend, that is usually observed in LMXBs due to a fluorescence 
mechanism driven by helium that principally results in the emission of NIII 4634/4640 \AA \, and CIII 4647/4650 \AA \, lines from the 
strongly irradiated face of the companion star. 
A possible feature consistent with the Bowen blend is present in our spectrum or even a P-Cygni profile for the weak He II 4686$\AA$\, line, 
but we consider them of too low S/N to led us to firm conclusions.


The best way to assess whether a LMXB is a UCXB is to measure its orbital period.
To this aim we analysed the optical light curves that we obtained with 
REM both in February 2015 (covering $ \sim 3.4 $ hours; Fig. \ref{fig_REM_feb}) and April 2015 (covering $ \sim 0.7 $ hours; Fig. \ref{lc_REM_apr}), but no evidence for a periodicity was found, only erratic variability.
This erratic variability is typical of LMXBs during outburst, and it hampers photometric studies, especially for UCXBs (\citealt{Nelemans2004}; \citealt{Baglio_4U0614_2014}).
We thus tried to constrain the orbital period of the system in an alternative way. Starting from the hypothesis of the ultra-compact nature of J180408, we evaluated its mass accretion rate $ \dot{M} $. To do this, we considered the {\it MAXI} and the {\it RXTE} light curves in the 2--10 and 2--20 keV band (respectively) and we integrated the counts of J180408 over the duration of the 2015 outburst. Since no other significant outburst
occurred
since the first detection of the system, we averaged our luminosity over the total time of the {\it MAXI} (first) and the {\it RXTE+MAXI} monitoring, obtaining $ \dot{M}= 4.0\times10^{-11} \, M_\odot/\rm yr$ and $ \dot{M}= 1.2\times10^{-11} \, M_\odot/\rm yr$, respectively (a Crab-like spectrum has been assumed for this conversion). We also point out that the short X-ray burst that took place in 2012 \citep{Chenevez2012} produces only a minimal change to the estimate of the mean accretion rate above. 
The transient behaviour in LMXBs is usually explained in terms of accretion disc instabilities (DIM model, \citealt{Lasota2008}). In order to exhibit outbursts a LMXB should have a mean mass accretion rate below a critical rate. For a He donor star (and orbital periods lower than 60 min) 
this implies a loose limit on the mass accretion rate to be below $\lsim 10^{-10} \, M_\odot/\rm yr$ \citep{Lasota2008}.
In addition, if we consider the evolutionary track of a He donor in a UCXB to match the mass accretion rate we worked out for J180408,
we can derive an estimate for the orbital period. Based on the tracks presented in \cite{Lasota2008} (their Fig. 3), we can infer 
an orbital period around 40 min for J180408. This is clearly an indirect inference and should receive an observational confirmation.

\begin{figure*}
\centering
\includegraphics[scale=0.55]{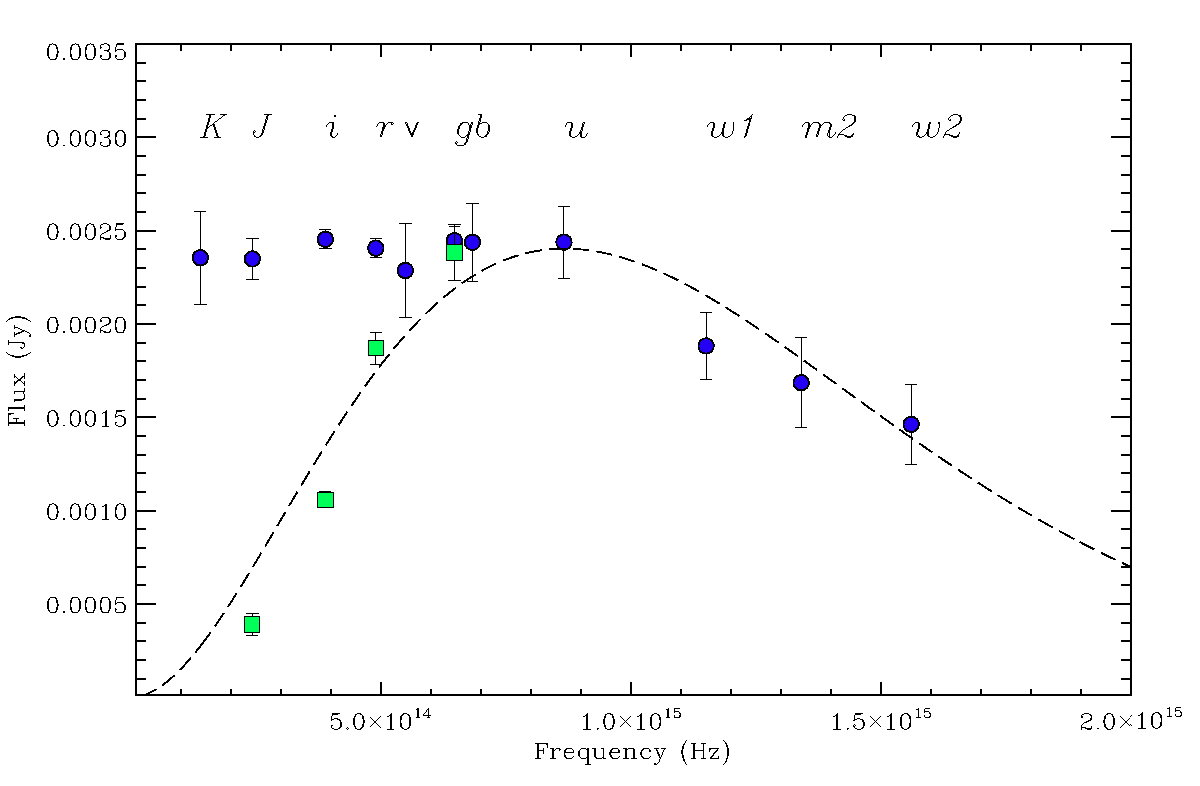}
\caption{NIR-Optical-UV spectral energy distributions of J180408 relative to the two epochs of Feb. 26, 2015 (MJD 57079) and Apr. 24, 2015 ({MJD 57136}). With green squares, the $ J $-, $ i $-, $ r $-, $ g $- bands observations obtained with the REM telescope on Apr. 24 are depicted; with blue dots, the points relative to the UVOT observations of Feb. 26 and the $ K $-, $ J $-, $ i $-, $ r $, $ g $- band REM observations taken on Feb. 26 are represented.}
\label{SED_feb}
\end{figure*}


\subsection{Spectral energy distribution}\label{SED}
The system J180408 was observed almost simultaneously (within 5 hrs) by REM (Tab. \ref{obs_log}) and \textit{Swift}/UVOT (Tab. \ref{tab_log_UVOT}) on Feb. 26, 2015 (MJD 57079). This allowed us to build a nearly simultaneous spectral energy distribution (SED) from the NIR to the UV wavelengths.
The REM optical (NIR) fluxes were obtained starting from the AB (Vega) magnitudes reported in Tab. \ref{obs_log}, corrected for the reddening due to the Galactic extinction in the direction of the target ($E(B - V)$=0.41, $A_{V}/E(B-V)=3.1$; \citealt{Schlafly2011}).
The UV and optical fluxes relative to the \textit{Swift}/UVOT observations were evaluated from the magnitudes of the target returned by HEASOFT (as described in Sec.\ref{sec_UVOT}) and reported in Tab. \ref{tab_log_UVOT}, after correcting for the interstellar absorption.

The same analysis has been applied to the second epoch of observations (Apr. 24, 2015 - MJD 57136), for which we possess the REM infrared and optical photometry (Sec. \ref{REM_sec}). In the $ K $- and $ H $-bands the target was too faint to be detected and \textit{Swift} did not observe the system on that occasion.

The NIR-optical-UV spectral energy distributions of the two different epochs are represented in Fig. \ref{SED_feb}. 
A rigid shift ($ 1.33\, \rm mJy $, that corresponds to a factor of $\sim 2$) has been applied to all the REM fluxes of the Feb. 26 epoch in order to be consistent with those of UVOT at the same wavelengths. 

We focused first on the Feb. 26 epoch and we looked at the highest frequency points ($V$ filter onward) in Fig. \ref{SED_feb} (blue dots).  
A fit with an accretion disc model (irradiated or not) does not provide an adequate description of the data. 
A fit with a black body model provides instead an acceptable fit ($\chi^2/\rm dof=1.3$, $30\%$ null hypothesis probability) 
with a surface temperature of $\sim 14,200 \, K$ (dashed line in Fig. \ref{SED_feb}). 
This black body emission may come from the hot spot in the stream-disc collision region, the companion star or the outer disc regions. 
The extrapolation of the black body spectrum largely misses the observations at lower frequency ($ r $, $ i $, $ J $, $ K $-bands).
This allowed us to hypothesise that the emission process dominating the red part of the SED is different from the one responsible for the blue part of the SED. The flat low-frequency spectrum is highly suggestive of a particle jet emitted from the central regions of the system. 

Due to the small frequency coverage of the second epoch SED (Apr. 24, 2015, the green squares in Fig. \ref{SED_feb}),
it was not possible to let the fit with a black body converge to reasonable values, since the peak of the black body was  unconstrained. 
However, the Apr. 24 SED seems to follow quite well the same black body as the Feb.  26 (the dashed line in Fig. \ref{SED_feb}). 
This indicates that the putative jet is a transient feature in the SED of J180408.

\begin{figure}
\centering
\includegraphics[scale=0.45]{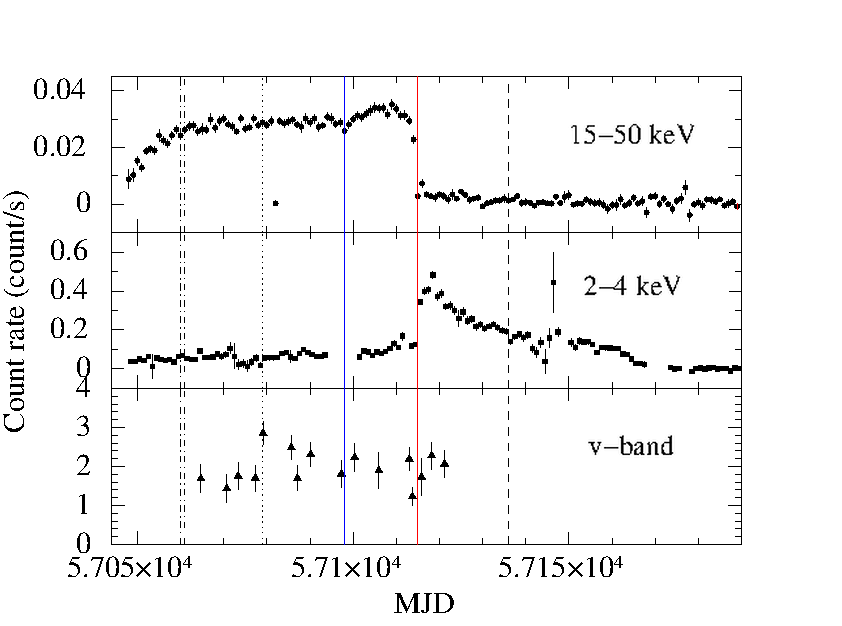}
\caption{From top to bottom, the monitoring of the system J180408 with the BAT instrument mounted on \textit{Swift} (15-50 keV), the {\it MAXI} camera (2-4 keV) and with UVOT ($ v $-band) during its outburst, starting from MJD 57048 (Jan. 26, 2015). The time of the detection of the radio jet \citep{Deller2015} and the time of the switch to a soft X-ray spectral state are highlighted with two solid lines (blue and red, respectively). With a dotted and a dashed lines the two epochs of our REM observations (Feb. 26 - Apr. 24, 2015) are marked. With two dot-dashed lines we finally indicated the times of the NTT photometric and polarimetric observations (Feb. 7-8, 2015).}
\label{Xray_monitoring}
\end{figure}

The presence of jets in LMXBs is usually associated with a hard X--ray spectrum. In the radio band steady jets observed to date 
during the hard X--ray states have a flat spectrum (e.g. \citealt{Fender01}). Above certain frequencies this flat spectral component should break to an optically thin spectrum. There are evidences both from black hole and neutron star X-ray binaries that this break occurs usually in the NIR (\citealt{Migliari10}; \citealt{Gandhi11}).
During its outburst J180408 was monitored by several all-sky instruments, allowing us to trace its spectral evolution (see Fig. \ref{Xray_monitoring}). As often happens in LMXB outbursts, J180408 started as a hard X--ray source and remained in the so-called hard state for about two months.
This is testified by the detection of J180408 with the \textit{Swift}/BAT hard X-ray telescope and the weak detection with the \textit{MAXI} low-energy monitoring instrument (see e.g. \citealt{Negoro2015} and Fig. \ref{Xray_monitoring}). On Apr. 3, 2015 (MJD 57115) a drop in the hard X-rays was observed \citep{Degenaar2015}, suggesting that the source transitioned to a soft X-ray spectral state (Fig. \ref{Xray_monitoring}). Consistently, the 
{\it MAXI} flux raised considerably.  

Our first SED has been taken in the hard X-ray state. As described above this SED shows a flat low-frequency tail, suggestive of a jet. 
The contribution of the jet seems to be not just restricted to the NIR spectrum, but appears to be important at least until the $ r $-band frequency is reached. The second SED has been taken when J180408 was in the soft X-ray state where a steady jet is expected to be quenched, thus explaining the different shapes of the two SEDs (e.g. \citealt{Russell2011}, \citealt{Munoz2014}).

In this scenario, the non-detection of linear polarisation when the source was in the hard X-ray state (Sec. \ref{Opt_pola}) seems to be puzzling.
In fact we would expect to detect some polarised emission at least at a few percent level in the optical-NIR bands \citep{Russell11}, 
if a jet were to be present. Our polarisation data were collected on Feb. 9 some weeks before our likely detection of a jet (Feb. 26). 
Even if we cannot build a wide band SED, our NTT data collected on Feb. 8 ($BVRi$) are consistent with the low-frequency tail of a black body, as in the case of the second epoch observations with REM, when the jet was not detectable in the SED.
We also note that at the time of the polarisation observations, the plateau that corresponds to the hard X-ray spectral state phase of the source (namely when the jet should first arise) was not yet completely reached (and the optical flux was a factor of $\sim 2$ lower). 
We can thus infer that the jet was probably not yet formed at the time of our polarimetric observations or, at least, it might not contribute 
substantially at optical ($BVRi$) frequencies.

\section{Conclusions}
In this work we performed a detailed NIR/optical and UV study of the LMXB 1RXS J180408.9-342058. Our principal results are summarised below.
\begin{itemize}
\item We observed the optical spectrum during outburst with the aim of identifying the nature of its companion star. The spectrum is rather featureless, with no H$\alpha$ and HeI emission lines. This is uncommon for bright LMXBs.
We only  marginally detected the He II 4684\AA\ emission line, suggesting the presence of helium. The lack of hydrogen lead us 
to hypothesise that the companion star could be a He white dwarf and J180408 be part of the UCXB class.
\item We searched our photometric data for an orbital period. No evidence for a periodicity was found. Integrating 
the long term X-ray light curve from all-sky monitoring instruments, we evaluated the mean mass accretion rate of the system. 
Based on the disc instability model \citep{Lasota2008} and the evolutionary track for a He white dwarf companion,  
we estimated the orbital period of J180408 to be $\sim 40$ min, that is typical of a UCXB. This inference clearly needs to be verified observationally.
\item We built the quasi-simultaneous SED from NIR to UV during the hard X-ray state. We interpret the SED as due to the emission of two components: a black body, and a steady jet that emits synchrotron radiation up to the $R$ band.
We also built the NIR/optical SED after the transition to a soft X-ray spectral state. As expected, during this epoch the jet seems to be quenched, leaving the black body contribution alone in the SED. 
\item We did not observe significant polarisation in the optical bands, despite J180408 being in the hard X-ray state. 
However, at the time of our polarimetric observation (weeks before our likely detection of the jet in the SED), the hard X-ray state was not  completely reached by the source. We conclude that at that time the jet probably was not yet formed, or it might contribute just at lower frequencies (i.e. the radio band).
\end{itemize}

\begin{acknowledgements}
We thank S. Covino and E. Palazzi for useful comments. MCB thanks George Hau and the La Silla staff for their support during her observing run.
TMD acknowledges support by the Spanish Ministerio de Econom\'{i}a y competitividad (MINECO) under grant AYA2013-42627
V.P.-A and V.C. acknowledges support by CONACyT research grant 151494 (M\'{e}xico). V.P.-A. acknowledges support from the CONACyT program for PhD studies.
\end{acknowledgements}

\addcontentsline{toc}{chapter}{bibliografia}


\end{document}